\documentclass[sigconf]{acmart}

\makeatletter
\def\@ACM@checkaffil{
    \if@ACM@instpresent\else
    \ClassWarningNoLine{\@classname}{No institution present for an affiliation}%
    \fi
    \if@ACM@citypresent\else
    \ClassWarningNoLine{\@classname}{No city present for an affiliation}%
    \fi
    \if@ACM@countrypresent\else
        \ClassWarningNoLine{\@classname}{No country present for an affiliation}%
    \fi
}
\makeatother

\usepackage[utf8]{inputenc}
\usepackage{subcaption}
\usepackage{multirow}

\title{WinkFuzz: model-based script synthesis for fuzzing}
\author{Zian Liu}
\email{102516622@student.swin.edu.au}
\affiliation{\institution{Swinburne University of Technology \& Data 61}}

\author{Chao Chen
}
\email{chao.chen@rmit.edu.au}
\affiliation{\institution{Royal Melbourne Institution of Technology}
            } 

\author{Muhammad Ejaz Ahmed
}
\email{ejaz.ahmed@data61.csiro.au}
\affiliation{\institution{Data 61, CSIRO}
            } 

\author{Jun Zhang
}
\email{junzhang@swin.edu.au}
\affiliation{\institution{Swinburne University of Technology}
            }   
\author{Dongxi Liu
}
\email{dongxi.liu@data61.csiro.au}
\affiliation{\institution{Data 61, CSIRO}
            }

\date{November 2022}
\newcommand{\mypara}[1]{\vspace{2pt}\noindent\textbf{{#1: }}}
\usepackage{xspace}
\begin{document}
\newcommand{\name}{WinkFuzz\xspace}
\newcommand{\Zian}[1]{\textcolor{orange}{[Zian - #1]}}

\begin{abstract}
Kernel fuzzing is important for finding critical kernel vulnerabilities. Close-source (e.g., Windows) operating system kernel fuzzing is even more challenging due to the lack of source code. Existing approaches fuzz the kernel by modeling syscall sequences from traces or static analysis of system codes. However, a common limitation is that they do not learn and mutate the syscall sequences to reach different kernel states, which can potentially result in more bugs or crashes. 

In this paper, we propose \name, an approach to learn and mutate traced syscall sequences in order to reach different kernel states. \name learns syscall dependencies from the trace, identifies potential syscalls in the trace that can have dependent subsequent syscalls, and applies the dependencies to insert more syscalls while preserving the dependencies into the trace. Then \name fuzzes the synthesized new syscall sequence to find system crashes.

We applied \name to four seed applications and found a total increase in syscall number of 70.8\%, with a success rate of 61\%, within three insert levels. The average time for tracing, dependency analysis, recovering model script, and synthesizing script was 600, 39, 34, and 129 seconds respectively. The instant fuzzing rate is 3742 syscall executions per second. However, the average fuzz efficiency dropped to 155 syscall executions per second when the initializing time, waiting time, and other factors were taken into account. We fuzzed each seed application for 24 seconds and, on average, obtained 12.25 crashes within that time frame.
\end{abstract}
\maketitle

\section{Introduction}
Vulnerability finding is of great importance in cyber security; it is the foundation of many other topics, such as fixing the vulnerability and learning from known vulnerabilities to find more similar ones. Among many existing vulnerability finding techniques, fuzzing is one of the most widely adopted approaches that has achieved remarkable results in real-world vulnerability finding, such as Google's OSS Fuzz project \cite{ossfuzz}. Fuzzing has also gained much popularity in research \cite{bohme2016coverage,stephens2016driller,johnson2004finding,wang2012improving}. Fuzzing approaches differ for different objects (e.g., binary code, source code, open-sourced kernel, close-sourced kernel, driver, etc.) due to the different object designs. In this paper, we focus on fuzzing the Windows system. System kernel vulnerabilities can be exploited to cause fatal attacks ranging from Blue Screen of Death (BSoD) to unauthorized accesses. Therefore, there are many existing researches focusing on this topic from both industry \cite{OSXFuzz,syzkaller,Trinity} and the academic field \cite{corina2017difuze,imf,jeong2019razzer,kim2020hfl,kim2017cab,li2016active,pan2017digtool,kafl,gauthier2011enhancing,weaver2015perf,garn2014eris}.

There already exist many works targeting kernel fuzzing. However, DIFUZE \cite{corina2017difuze}, HFL \cite{kim2020hfl}, seL4 \cite{klein2009sel4}, and \cite{klein2014comprehensive} require Android or Linux source code for analysis, which is generally unavailable for closed-source systems. Some other works focus on specific attack surfaces. For example, \cite{BrokenType} focuses on font-related APIs, and \cite{kim2017cab} focuses on fuzzing the IOCTL interface, improving bug-finding efficiency by constructing pre-contexts using real programs and prioritizing paths that are likely to have bugs. Ioctl Fuzz \cite{Ioctl} also targets the IOCTL interface by hooking a \texttt{NtDeviceIoControlFile} API call and randomly mutating its argument values. Some other kernel fuzzers focus on one or several types of kernel bugs. For example, \cite{jeong2019razzer} focuses on kernel race bugs, and \cite{pan2017digtool} targets four kinds of kernel bugs (i.e., UNPROBE, TOCTTOU, UAF, and OOB) by using hypervisor technologies. Some kernel fuzzers are embedded with specific knowledge to find bugs; for example, Trinity \cite{Trinity}, iknowthis \cite{iknowthis}, and sysfuzz \cite{sysfuzz} are Linux kernel fuzzers built with hard-coded rules and grammars. \cite{KernelFuzzer,kafl,blackhat} require users to provide knowledge in order to synthesize harness code for fuzzing. Therefore, an automated kernel fuzzing approach that targets the general attack surface of kernel space is in demand. 

IMF \cite{imf}, Moonshine \cite{moonshine}, NtFuzz \cite{ntfuzz}, and \cite{model_fuzz} are automated kernel fuzzing approaches. IMF \cite{imf}, Moonshine \cite{moonshine}, and \cite{model_fuzz} trace syscalls from program execution and analyze dependencies from the trace. NtFuzz \cite{ntfuzz} executes the original programs for fuzzing, while \cite{model_fuzz}, IMF \cite{imf}, and Moonshine \cite{moonshine} generate valid model programs that preserve the syscall dependencies. The executed program can ultimately invoke syscalls with legitimate ordering and dependencies. The invoked syscalls are the only triggers that interact with the kernel to make the kernel reach deep kernel states. However, as we manually inspected the syscall trace, we found that the traced syscalls can be further mutated to make the kernel reach new states. For example, at some point in the trace, we observe that syscall2 takes the output value from syscall1 as the input argument. Then, at some point later in the trace, we observe syscall1 is invoked again, but this time without a subsequent dependent syscall2. Hence, we can insert an invocation of syscall2 after syscall1 and adhere to the argument dependency between them to synthesize a new syscall sequence. 

Syzgen \cite{syzgen} generates new syscall sequences by symbolically analyzing macOS driver binaries. Its target are macOS driver syscalls that are defined in the driver binaries, which it can symbolically analyze. On the other hand, IMF \cite{imf} and Moonshine \cite{moonshine} target Linux kernel and require kernel source code. For Windows kernel, NtFuzz \cite{ntfuzz} and \cite{model_fuzz} target syscalls that are not defined but only used in the system binaries. 
When calling these syscalls, the system directly switches to kernel mode from user mode. According to \cite{windows_kernel_func2}, these syscalls are handled by the syscall handler in kernel mode. The syscall handler further resolves kernel routines (functions, e.g., Nt!NtCreateFile) to provide the system service required by the syscall (e.g., ntdll!NtCreateFile) \cite{windows_kernel_func1}. These kernel routines do not exist in the kernel binaries, and can only be accessed via the de facto Windows system debugging tool, WinDbg, which renders the methods in Syzgen \cite{syzgen} ineffective. 

Even though works targeting user-space applications have a different research object than kernel fuzzing, some approaches still render inspiring ideas and techniques for us. Langfuzz \cite{langfuzzing} generates novel test cases by learning grammars from the existing ones that can lead to crashes or bugs. Randoop \cite{Randoop} uses feedback information to guide input mutation. Winnie \cite{winnie} targets Windows-based applications. It dynamically traces the application during execution and generates harness code based on the trace. 
Even though this genre of approach cannot be directly applied to kernel fuzzing, we used their idea to mutate test cases by learning from existing valid test cases (syscall logs).



To overcome the mentioned problems, we combine the design from both kernel fuzzing and user-space application fuzzing to generate and mutate syscall sequences from the logged trace. Our approach learns the syscall dependencies from the syscall trace and inserts them into the traced syscall sequence in order to trigger different deep kernel states. In this paper, we propose \name, a kernel fuzzing approach that generates and mutates syscall sequences from the trace. Firstly, \name utilizes NTfuzz's fuzzing module to log syscall traces. Secondly, \name analyzes the dependency relations between the traced syscalls. Thirdly, \name generates a model script that recovers the traced syscalls. Fourthly, it utilizes the argument type information (inferred by NTfuzz) and our inferred dependency relations to synthesize new syscall sequences. Finally, \name iteratively executes the synthesized syscall sequences and mutates the syscall arguments for fuzzing.

We evaluated \name on four seed applications and found that it increased the total number of syscalls by 70.8\%, with a success rate of 61\%. This means that \name can synthesize new syscall sequences that lead to different kernel states. We also evaluated the time efficiency of offline processing time and the online fuzzing rate. The average times for tracing, dependency analysis, recovering the model script, and synthesizing the script were 600, 39, 34, and 129 seconds, respectively. The instant fuzzing rate was 3742 syscall executions per second, but when we average extra time cost including the initializing time, waiting time, and other factors, the average fuzz efficiency dropped to 155 syscall executions per second. We fuzzed each seed application for 24 seconds and, on average, we obtained 12.25 crashes within that time frame.

Note that we use the term \textit{syscall} and \textit{system call} interchangeably. To conclude, we make the following contribution:
\begin{itemize}
    \item We expanded NTfuzz's fuzzing module to enable dynamic syscall tracing, which enables us to record each syscall with their input arguments and outputs.
    \item To mutate the contents in the system's memory address, we implemented the mutation module in \name to interact with NtFuzz's mutation module and utilize it for syscall argument mutation.
    \item We implemented a tool called \name that can statically analyze Windows syscall dependencies among each other, based on the traced syscalls automatically. \name further applies the learned syscall dependencies to generate more valid syscalls to reach different kernel states. Our tool is available at \url{https://anonymous.4open.science/r/Wink-fuzz-0067/}. After this paper is accepted, the tool and dataset will be published. 
    
\end{itemize}
\section{Overview}
\subsection{The Motivating Example}
\begin{figure*}[]
\centering
\includegraphics[width=0.6\textwidth]{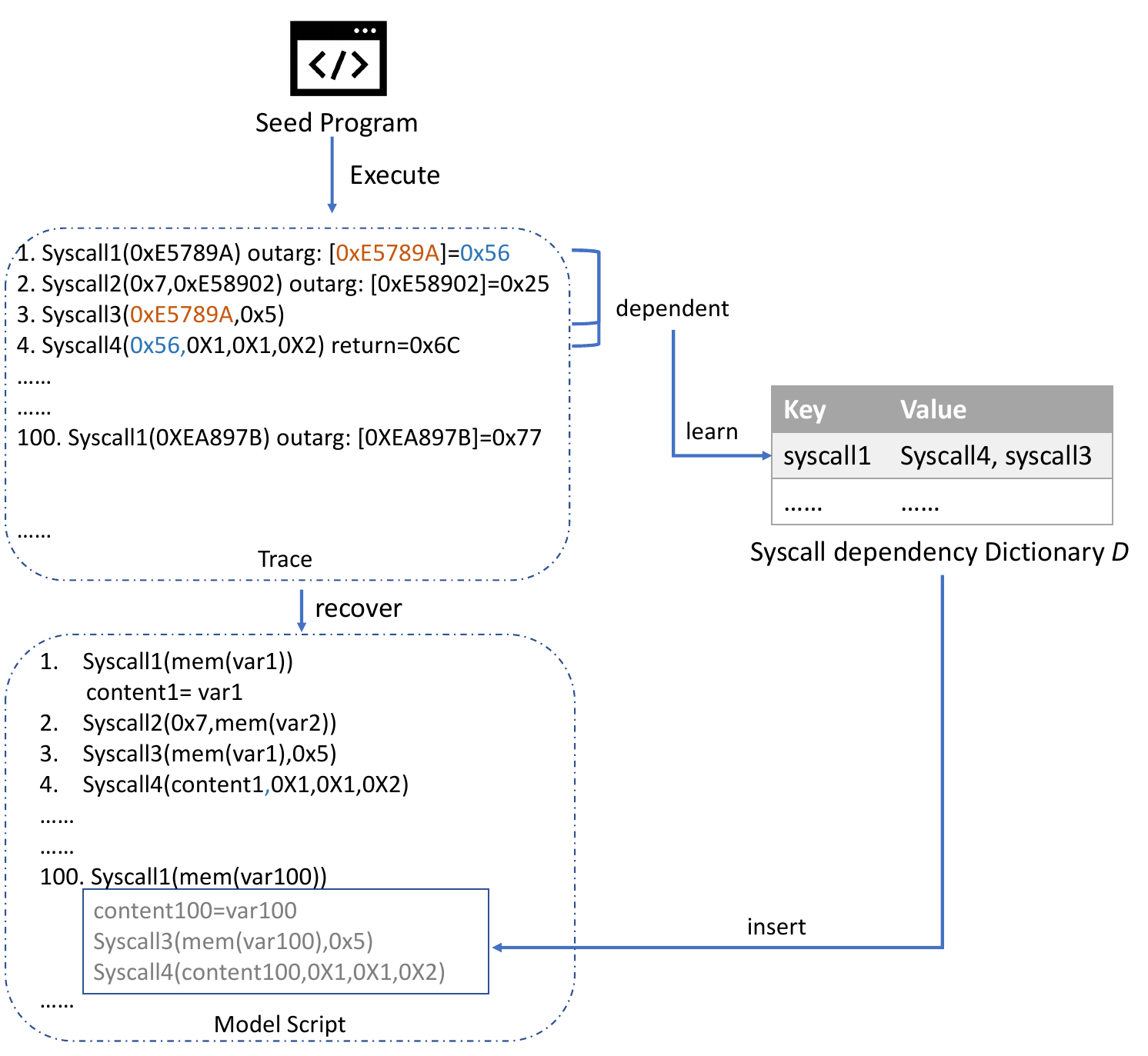}
\caption{An example of synthesizing model script for fuzzing.}
\label{fig:motive}
\end{figure*}

\autoref{fig:motive} demonstrates an example of our approach to synthesize a script for fuzzing. Firstly, we select a seed program and execute it. During the execution, we log the system calls invoked due to the seed program, along with their input arguments, output arguments, and return values. As shown in the \textit{Trace} rectangle in the figure, we record input arguments for each system call. Syscall$i \quad (i \in N$) represents the invoked system call name. The values within the bracket separated by a comma are the arguments for the system call. \texttt{outarg:} records the output argument (which is a pointer to a memory address), along with the output contents that are received in the memory address. \texttt{return:} records the system call's return value. If the system call does not have an output argument or a positive return value, then we omit the \texttt{outarg:} or the \texttt{return:} in the figure as only positive return values can be dependent values.
In the example, we observe that, at line 1, after syscall1 completes, an output value 0x56 is returned to the memory address pointed to by its first argument, 0xE5789A. Later, at line 3, syscall3 reuses the memory address 0xE5789A as its argument. Also, syscall4 uses the content (0x56) in that memory address as its argument. Therefore, syscall3 and syscall4 both depend on syscall1. These dependencies are first used to recover a model script as \texttt{Model Script} in the figure, in lines 1, 3, and 4. We preserve the dependency between syscall1, syscall3, and syscall4 by using the variables \texttt{var1} and \texttt{content1} to keep the dependencies. Then, we use this learnt dependency again to synthesize a new script. Specifically, we use the argument from syscall1 at line 100 as the inserted syscall3's argument, which is a pointer to a memory address. And we use the content within the pointer as an argument for the inserted syscall4. For the non-dependent arguments in syscall3 and syscall4, we just copy their values from the trace (e.g., the second argument of syscall3 and the last three arguments of syscall4).

Existing model-based kernel fuzzing methods stop at the modeling script step and use the modeled script for fuzzing. However, as we have described, the modeled script can be further mutated to obtain new syscall sequences that can reach different kernel states. Existing work tries to explore more dependencies by analyzing the kernel source code or symbolically executing the system binaries, which contain syscall definitions. However, this is unpractical for our target Windows syscalls, as these syscalls are neither open source nor defined in the system binaries. Alternatively, we learn the dependencies from the trace and apply them for trace mutation, as shown in the example. 

\subsection{Workflow}
\name contains five steps: (1) dynamic syscall tracing, (2) syscall dependency analysis, (3) model script recovery, (4) new script synthesizing, and (5) fuzzing. We briefly introduce each step here.

\mypara{Dynamic Syscall Tracing} In the first step, we select a seed application, execute it, and log the invoked system calls during the execution. We expand the fuzzing module from NtFuzz to dynamically trace the system calls invoked, along with their arguments, and return results. We prepare a script file for each seed application, which opens up the target application and interacts with it, by mimicking a human user's behavior (like clicking the mouse, zooming in the window, etc.).

\mypara{Syscall Dependency Analysis} In the second step, \name analyzes the recorded trace offline to determine the interdependency between syscalls. The dependencies are inferred by matching the input and output argument values among different syscalls.

\mypara{Model Script Recovery}
In this step, \name recovers the syscall sequence according to the traced logs. The ordering and dependencies of the syscalls strictly preserve those in the log. Since there are too many types of structures used by the syscalls, \name first defines the structures by declaring each field of the structures. Then \name prepares the syscall arguments with the traced values. The recovered model trace can still encounter severe errors that hinder the complete execution. To execute all the syscalls in the model script, we manually identify and rectify any erroneous syscall arguments.

\mypara{Script Synthesizing} In this step, \name synthesizes new syscall sequences by inserting syscalls with the learnt dependencies. To insert syscalls, \name firstly analyzes the trace and finds insertable sites, which are the syscalls in the trace that could be followed by subsequent dependent syscalls but are not in the trace. Then \name inserts the dependent syscalls after those syscalls while preserving the learnt syscall dependencies. The insertion can be iterative (i.e., further inserting syscalls that depend on the newly selected syscalls). In order for the synthesized script to be executed without any hangs or exits, we manually observe and rectify any syscall arguments that cause severe errors.

\mypara{Fuzzing} In this step, \name fuzzes the kernel by executing the synthesized script repeatedly. During the execution, the syscall arguments in the synthesized script are mutated. Even though the syscall argument can involve many different data types ranging from integers, strings, to structures, when invoking the syscall, they can be categorized into two types: constant values and pointers pointing to various object types. For constant values, \name mutates them on its own. For pointer type arguments, \name needs to mutate the values in the memory addresses recursively. For simplicity, since NtFuzz's fuzzing module has the functionality to recursively mutate memory addresses, \name interacts with NtFuzz's fuzzing module to mutate the content pointed to by pointers.
\section{Methodology}
\label{sec:meth}

\begin{figure*}[!ht]
\centering
\includegraphics[width=\textwidth]{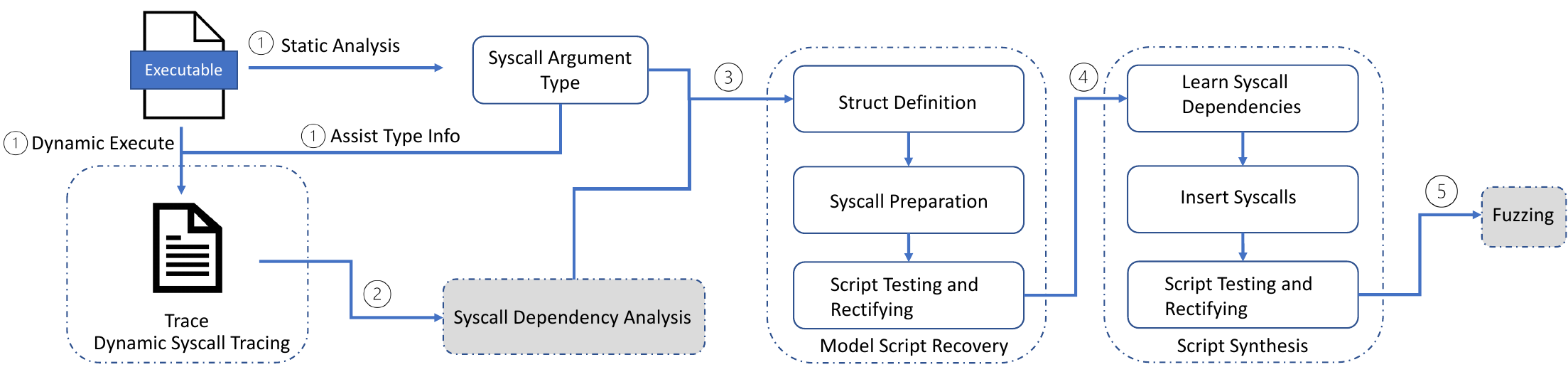}
\caption{Methodology of \name. We firstly use NTfuzz to output argument type information for each syscall as this output will be used in our methodology. Then \name consists of three steps: 1) Dynamic Syscall Tracing, 2) Syscall Dependence Analysis, and 3) Script Generation.}
\label{fig:meth}
\end{figure*}

We introduce the details of our approach in this section. The methodology consists of five steps: 1) dynamic syscall tracing, 2) syscall dependence analysis, 3) model script recovery, 4) new script synthesizing, and 5) fuzzing, as shown in \autoref{fig:meth}.

\subsection{Dynamic Syscall Tracing}
\label{sec:meth:trace}
The Windows system kernel utilizes more than 1600 syscalls to implement internal functionalities. Fuzzing these syscalls is of great interest for kernel fuzzing since they can cause fatal kernel panics. However, these syscalls are undocumented, which results in two major problems: 1) The argument types are unknown. Without argument type information, it is not possible trace syscall with correct argument values. 2) The valid syscall ordering is unknown. When the argument type is known, the syscall ordering is also of great importance. Since there are too many syscalls in Windows, blindly calling them without knowing the valid ordering would lead to many errors and thus not be able to reach deep kernel status.

Luckily, a tool called NTfuzz, which was proposed recently, is able to solve these two problems. NTfuzz can statically analyze syscall types by using symbolic execution across system binaries. Therefore, we directly utilize NTfuzz to statically recover Windows syscall argument type information. Secondly, we expand the fuzzing module of NTfuzz, as described in \autoref{sec:meth:expand}, to dynamically trace the invoked syscalls with their arguments and return values. Since the behaviours of arbitrary applications on the Windows platform can ultimately invoke Windows syscalls, the traced logs must have valid ordered syscalls.

\subsubsection{Expanding Fuzzing Module of Ntfuzz}
\label{sec:meth:expand}
NtFuzz fuzzes syscalls by intercepting each one prior to execution and mutating the argument fields. To intercept the syscalls, NtFuzz installs an agent driver in the system to identify the target syscalls and intercept them. When the script file is run, the location of the target seed program in the system is passed to NtFuzz's fuzzing module. 
As a result, only the system calls invoked due to the target applications are logged. To mimic the interaction with the applications as humans, each seed program is prepared with a script file. By executing the script file, a set of pre-written interactions with the applications will be performed, such as clicking the buttons, maximizing/minimizing the window, or typing on the keyboard.

For simplicity, \name utilizes this agent driver from NtFuzz to log target system calls. We have expanded NtFuzz's fuzzing module so that it does not mutate system call arguments, but instead logs those arguments after intercepting the system calls. The expanded fuzzing module also records the system call return value after the system call has finished.

\subsection{Syscall Dependency Analysis}
\label{sec:meth:Dependency}
\begin{figure*}[!h]
\centering
\includegraphics[width=0.6\textwidth]{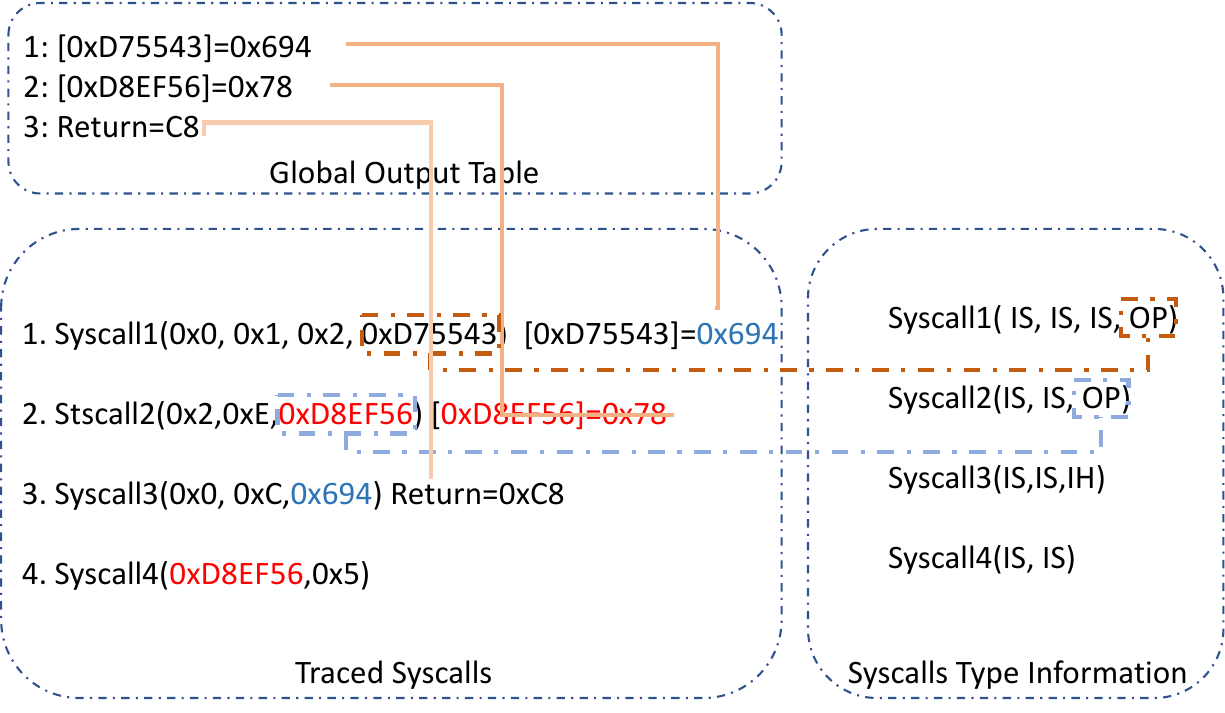}
\caption{
An example is shown to demonstrate how \name finds inter-syscall dependencies by checking syscall type information and storing output values to a global output table $\mathcal{T}$. The type information of each argument consists of two alphabetical characters: the first character \textit{I} represents that the current argument is an input, while \textit{O} indicates an output argument. The second character includes \textit{S,H,P,A,F}, which stands for scalar (i.e., constant), handle, pointer, array, and function pointer types, respectively. The \textit{[]} symbol represents the value stored in the memory address that is in between the brackets.}
\label{fig:dependent_example}
\end{figure*}
Windows system calls can rely on values from previously executed system calls (i.e., either return values or argument values that receive outputs after the system call execution). The dependent value may be a variable rather than a constant, such as a handle value or memory address. Therefore, despite correctly reserving the system calls with valid ordering, it is vital to discover such dependencies among system calls in order to generate a working model system call sequence.

In \autoref{sec:meth:trace}, we have already inferred the syscall argument type information by using NtFuzz. The type information refers to the type of each argument. If the type involves structures, the type information refers to the structure template (i.e., the type and offset of each field). To find the dependency relations, \name scans the trace from the beginning. For each syscall, \name records its return value or output argument values into a global output table $\mathcal{T}$, ordered based on the order of scan in the log. We decide whether to record a value to $\mathcal{T}$ by checking its syscall type information. If the argument type is an output or when we encounter a return value, we add it to $\mathcal{T}$. When we scan the following syscalls, we iteratively check their arguments. If the argument $arg_i$ is a dependent type (e.g., handle type), or if the value of $arg_i$ is likely to be a memory address (i.e., hexadecimal with at least six digits), we check each return or output value in $\mathcal{T}$ in the reverse order. If $arg_i$ matches any value in $\mathcal{T}$, that implies we have found a dependency relation.

For example, \autoref{fig:dependent_example} demonstrates a toy example. Assume we have already used NtFuzz to infer syscall argument information. Now we start scanning the trace to find dependencies. The fourth argument of syscall1 is an output pointer argument with value 0xD75543. Since the trace records the fourth argument as a pointer pointing to a scalar (constant) type value, the argument value (0xD75543) and the value in that memory address (0x694) are also recorded in $\mathcal{T}$. The same step is done for syscall2's third argument (0xD8EF56). Then for syscall3, we add this return value to $\mathcal{T}$ since this syscall has a positive return value of 0xC8. Zero or negative return values are not recorded in $\mathcal{T}$ because zero can either indicate the syscall executed successfully or unsuccessfully, and negative values are Windows error codes, and only positive return values can contain dependent information such as the memory address or the handle. Then we continue scanning the trace. We then scan syscall4 in line 4. Since the first argument is a scalar (constant) type and the value looks like a memory address, we check whether this value matches any value in $\mathcal{T}$ backwardly, from the end to the beginning of the table. That is, we firstly check the third row \texttt{3: Return=C8}, then the second row \texttt{2: [0xD8EF56]=0x78}, then we check the first row \texttt{1: [0xD75543]=0x694}. It turns out the first argument of syscall4 matches the second argument of syscall2, indicating a data dependency between syscall2 and syscall4. As a result, \name generates a file recording the argument dependencies among all syscalls.

\subsection{Model Syscall Recovery}
In this step, \name automatically generates a model script that preserves the execution order and dependencies of the syscalls in the logged trace.

To generate a model script, there are two steps: 1) structure definition, and 2) syscall preparation. This is because syscalls take various structures as arguments. There are too many different types of structures used by the trace, so it is necessary to correctly define these structures before using them as arguments in the syscall. 

\mypara{Structure Definition}\label{sec:meth:struct_define}
Structure definition is built upon the analyzed argument type information by NTfuzz in \autoref{sec:meth:trace}. Note that if one structure contains structures recursively, we need to define all of the structures.

\mypara{Syscall Preparation}
In this substep, \name automatically generates script codes that initialize the argument values and fill in the correct value for each field of the structure for the syscalls, as recorded in the trace. Then \name prepares script code that invokes the syscalls. The generated syscalls should strictly preserve the recorded argument value, the syscall ordering, and dependencies in the trace. Specifically, to preserve the dependencies, if the argument is a \texttt{handle} type or is likely to be a memory address, \name would trace backwards to the previous syscall that is the source of this \texttt{handle} or memory value, and propagate the value from the source syscall to the current destination syscall(s).

\mypara{Model Script Testing and Rectifying}
\label{sec:meth:model_test_rect}
However, our dependent NtFuzz may introduce errors because the same system call may exist at multiple sites in system binaries. NtFuzz may infer different argument information at different sites for the same system call. The difference may come from 1) the inference errors (reported to successfully infer 69\% of the system call arguments in \cite{ntfuzz}, or 2) multiple usages of the system calls, as the same Windows system call can have different templates (i.e., having different value types for the same argument). For generality, NtFuzz votes for the majority, which can ignore the inference errors and the legitimate minority cases. However, the traced arguments are based on the voted majority template.

Therefore, the modeled script can cause severe errors when the syscall in the trace was not used as the majority use case. These errors can hinder the model script execution (i.e., the remaining syscalls in the script cannot be executed). The incorrect argument can be either an input argument or an output argument (that receives an output value after the syscall completes). The erroneous input argument causes the error because it provides an invalid value, while the erroneous output argument causes an error because the output memory is not large enough to receive the outputs.

Unfortunately, there are currently two challenges in this step: 1) Knowing the specific usage of each syscall in the trace; and 2) Locating the erroneous argument automatically, as the error is not a traditional error that can be debugged by the debugger. Therefore, we can only find them manually. Specifically, we narrow down the potential erroneous locations by commenting part of the model script, and use a binary search algorithm to boost the process. Once we locate the root cause argument, if the error is due to an erroneous input argument, we discard the syscall because we cannot know the correct input value to replace it. If the error is due to an erroneous output argument, we allocate a much larger memory to replace the old argument.

\subsection{New Script synthesizing}
Up until now, \name has generated a model script that recovers the syscalls according to the trace. In this section, \name further inserts syscalls in the model script to synthesize a new syscall script. \name selects existing syscalls that can have, but do not have, subsequent dependent syscalls as insertable sites. Then, \name inserts dependent syscalls after the insertable sites. Therefore, the kernel can reach new states that cannot be reached by the modeled script. To synthesize a new script, \name 1) learns the syscall dependencies, 2) searches for insertable sites in the trace and inserts possible syscalls after them.

\subsubsection{Learn Syscall Dependencies}
\label{sec:meth:learn}
Based on the output file that records the argument dependencies among syscalls in \autoref{sec:meth:Dependency}, \name further generates a dictionary $\mathcal{D}$ recording the legitimate usage between every two directly dependent syscalls. The keys in the dictionary are the source syscalls from which the dependent argument propagates. The values of each key in the dictionary are the destination syscalls to which the arguments directly propagate. \autoref{fig:synthesize} shows a further example of \autoref{fig:dependent_example}. From the logged trace, \name learns that syscall3 depends on syscall1, syscall4 depends on syscall2, etc. and records the learnt dependencies into a dictionary $\mathcal{D}$. Note that the specific information of the dependent argument (e.g., Syscall1's last argument is a memory address, and syscall3's last argument depends on the value in that memory address; syscall4's first argument depends on syscall2's last argument) is also recorded in the dictionary, but for simplicity, they are not shown in the figure.

However, not all of the learned syscall sequences in the dictionary $\mathcal{D}$ can execute successfully. Even the modeled script cannot guarantee a 100\% success rate, as seen in \autoref{sec:meth:model_test_rect}. Therefore, to reduce the failure rate when inserting new syscalls, \name observes the return value from the trace and only learns from the successfully executed syscalls, only inserting these successful syscalls.

\subsubsection{Insert Syscalls}
\label{sec:meth:insert}
\begin{figure}[!h]
\centering
\includegraphics[width=0.5\textwidth]{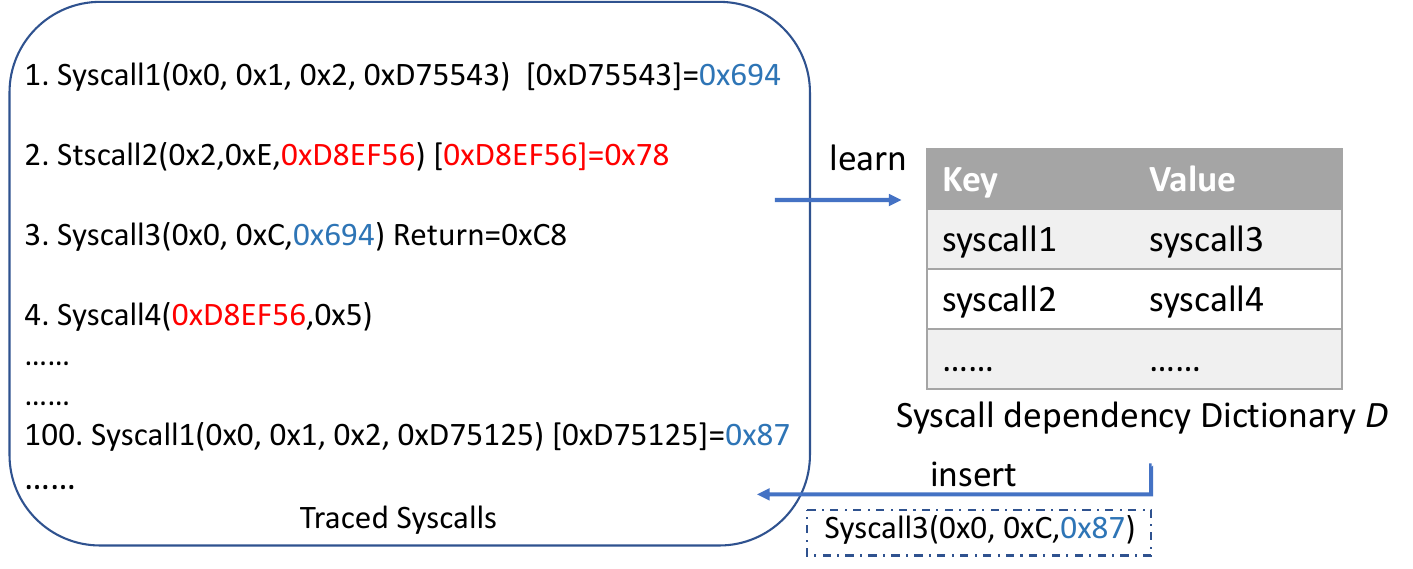}
\caption{An example of learning syscall dependency dictionary $\mathcal{D}$ is shown. The dependencies are learnt from the traced syscalls on the left hand side and inserted into the dictionary $\mathcal{D}$ on the right hand side. The learnt syscall dependencies are then used to insert new syscalls into the trace.}
\label{fig:synthesize}
\end{figure}

\name analyzes the logged trace and finds insertable syscalls ${sys\_insert}$ that can have subsequent dependent syscalls but not have them. Note that if we learn the syscall dependency from $syscall_a$ at some line $A$, and find that $syscall_a$ is called again at another line $B$, with the same following dependent syscalls as $A$, then $B$ is not an insertable site because inserting at $B$ cannot reach new kernel states. \name inserts dependent syscalls after those insertable sites, with well-preserved dependencies. We can recursively insert syscalls that depend on the just-inserted syscalls. This forms multiple levels of the inserted syscalls. 

An example is shown in \autoref{fig:synthesize}: \name learns that syscall3 depends on syscall1, and finds there is no subsequent syscall3 depending on syscall1 at line 100. Therefore, \name inserts a syscall3 with its last argument equal to the value (0x87) in the memory address represented by the last argument of syscall1 at line 100 (0xD75125). The inserted syscall in the example only contains one level of the dependent syscall. It is possible to insert more than one level of syscalls, by further inserting syscalls that depend on the just inserted syscall3. Based on the number of levels of the insertion, we call them $Ln$ (Level n) inserted syscalls, where $n$ denotes the index of the level. 

\mypara{Synthesized Script Testing and Rectifying}

Like \autoref{sec:meth:model_test_rect}, the inserted system calls can also result in severe errors such as hanging or exiting. These errors are beyond the capability of the debuggers, so no accurate error information can be provided for us to locate the erroneous locations. Therefore, we adopt the same method as \autoref{sec:meth:model_test_rect} to manually narrow down the suspicious erroneous location. We discard the system calls having erroneous input arguments because we do not know the correct input, and we enlarge the memory size for the erroneous output arguments.

\subsection{Fuzzing}
\name repeatedly executes the synthesized script and randomly mutates each system call's arguments in order to fuzz the kernel. However, according to \cite{syzkaller,model_fuzz}, \texttt{handle} type should not be mutated as mutating the \texttt{handle} type can greatly affect the success rate of system calls. Thus, \name also does not mutate \texttt{handle} type. Even though there are many types of input data (i.e., integer, string, handle, struct, array, pointer), in the system call argument's aspect, there are only two types: constant or pointer. The constant can be an integer or a handle, while the pointer points to somewhere in the memory. Therefore, for constant argument, \name adopts the same strategy to mutate them as NtFuzz: 1) bit flipping, 2) arithmetic mutation, 3) replacing with extreme or boundary values such as 0 or INT\_MAX, 4) replacing with random value. For pointer arguments, \name utilizes the fuzzing module in NtFuzz for mutation. Therefore, \name can inherit NtFuzz's type-aware mutation feature rather than blindly mutating. \name sets a threshold of executed system call numbers to trigger the mutation. \name also sets a mutation rate which represents the probability of mutation rather than mutating all the time, in order for fewer system call failures and thus deeper kernel states, as \cite{ntfuzz}.

Ntfuzz's fuzzing module is a driver installed in the Windows system and accessible through our synthesized script. The fuzzing module provides functionality to mutate the value in any memory address, or even recursively mutate contents if the argument includes recursive memory pointers. For example, if we would like to mutate the syscall argument which is a pointer type pointing to a structure, the structure containing a pointer pointing to another structure in the memory, the fuzzing module is able to mutate all the fields in all structures.
\section{Experiments}

\mypara{Testbed}
We conducted the experiment on a Windows 10 17134.1 installed in Virtualbox, as kernel fuzzing can lead to system crash and reboot. Therefore, it is better to test in a virtual environment, as cited in \cite{ntfuzz}. The host machine was an Intel NUC8i5BEH with 16 GB of RAM and an i5-8259U CPU with a 2.30GHz clock speed.

\mypara{Seed Application}

The seed application is used to trace the invoked system calls during its execution. Since arbitrary applications can ultimately invoke a valid system call sequence, four seed applications were randomly selected, including AdapterWatch v1.02, DxDiag, WordPad, and NotePad. \autoref{tab:seed_app} shows the number of traced system calls for each seed application. AdapterWatch contains the most system calls (16895) while NotePad has the minimum number of system calls (7487).

\begin{table}[!ht]
\centering
\caption{}
\label{tab:seed_app}
\footnotesize
\begin{tabular}{c|c}
\hline
  Application  & Traced Syscall (\#)\\
 \hline
 AdapterWatch&16895\\
 DxDiag&17579\\
 WordPad&41203\\
 NotePad&7487\\
\hline
\end{tabular}
\end{table}

Our research questions (RQs) are:
\begin{itemize}
    \item \textbf{RQ1.} How many new syscalls can be inserted and can they be executed correctly?
    \item \textbf{RQ2.} What is the efficiency of \name?
    \item \textbf{RQ3.} Can \name find system crashes during fuzzing?
\end{itemize}

\subsection{Inserted Syscall Number and Success Rate}
\begin{table*}[!ht]
\centering
\caption{Inserted syscall numbers, their ratio (compared to the traced syscall number), and success rate at each insert level $Ln$.}
\label{tab:synthesize}
\footnotesize
\begin{tabular}{c|ccc|ccc|ccc|ccc}
\hline
  Application & L1 syscall (\#)&inc. &success& L2 syscall (\#)&inc.& success & L3 syscall (\#) &inc.&success& total syscalls (\#) &total inc.& total success\\
 \hline
 AdaptWatch &4609 &27.3\% &68.8\%&2739 &16.2\% &60.8\% &4123 &24.4\% &54.9\% &11480 &67.9\% &61.8\%\\
 DxDiag &2561 &14.6\% &66.6\% &1733 &9.86\% &39.2\% &420 &2.38\% &0\% &4714 &26.8\% &50.6\%\\
 WordPad &19208 &46.6\%  &50.0\% &16442 &39.9\% &74.9\% &23065 &55.9\% &67.2\% &57815 &140.3\% &64.2\%\\
 NotePad &1281 &17.1\% &66.7\% &1481 &19.8\% &71.7\% &857 &11.4\% &67.4\% &3619 &48.3\% &67.4\%\\
\hline
Avg. & 6915 &26.4\% &63.0\% &5599 &21.4\% &61.7\% &7116 &23.5\% &47.4\% &19407 &70.8\% &61\%\\
\hline
\end{tabular}
\end{table*}

As we mentioned in \autoref{sec:meth:insert}, one can recursively insert dependent syscalls to synthesize syscall scripts. We use the symbol $Ln$ to denote the level of the inserted syscalls, where $L$ denotes the level and $n$ represents the number. We choose the number of levels ($n$) to be between 1 and 3, since with $n$=3, the inserted syscall number already grows enormous (i.e., the maximum total increased syscall can reach 140.3\%). Note that the inserted syscalls in each level $Ln$ do not overlap. Following \cite{model_fuzz}, we observe the return value of the newly inserted syscalls to measure the success rate. Since \name only learns and inserts the syscalls in the trace that successfully returned, we compare the inserted syscall's execution output (return value) with the original output in the trace. If they are the same, the newly inserted syscall is considered to have been successfully executed. We measure the percentage of the successfully executed inserted syscalls as the success rate.

\autoref{tab:synthesize} shows the results of the number of inserted syscalls and their success rate in each level. Wordpad has the most inserted syscall numbers in each level and in total (i.e., 19208 syscalls in $L1$, 16442 syscalls in $L2$, 23065 syscalls in $L3$, and 57185 syscalls in total). For $L1$ and $L2$, NotePad has the least inserted syscall numbers, with 1281 syscalls and 1481 syscalls, respectively. For $L3$, Dxdiag has the least inserted number (420 syscalls). In terms of inserted syscalls success rate, all four seed applications achieved a success rate above 50\%, with AdaptWatch having the most percentage (68.8\%) in $L1$. In $L2$, all applications except DxDiag achieved success rates above 60\%. DxDiag had the least success rate (39.2\%). In $L3$, again, all applications except Dxdiag achieved success rates higher than 54\%. DxDiag had a zero success rate this time. To understand why Dxdiag had much less success rate than the other applications, we manually analyzed the result and found that even though there were 420 $L3$ inserted syscalls, there were only 2 unique syscalls in $L3$. And these 2 unique syscalls could not execute successfully. Therefore, no matter how many times they repeated, the success rate remained 0\%.

As we described in \autoref{sec:meth:learn}, \name only learns and inserts successfully executed syscalls from the trace. However, this still cannot guarantee that all the inserted syscalls can be executed successfully. This is because: 1) It is possible that the inferred majority argument type by NtFuzz is not applicable at the inserted site. 2) The Ntfuzz type reference module only achieves a 69\% accuracy as reported in \cite{ntfuzz}. To insert as many new syscalls that are absent in the trace, our best practice is to conservatively insert all the syscalls as described in \autoref{sec:meth:learn}.

\begin{figure}[!ht]
\centering
\includegraphics[width=0.45\textwidth]{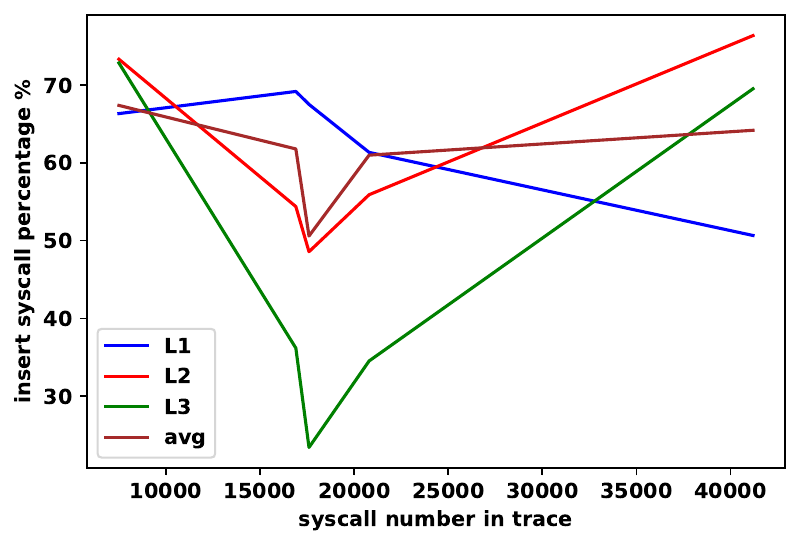}
\caption{Inserted syscall number percentage versus the syscall number in the trace for each level L1, L2, and L3.}
\label{fig:inserted_syscall}
\end{figure}
To understand whether the syscall number in the trace can impact the inserted syscall numbers in each level, we measure all of them. \autoref{fig:inserted_syscall} shows the relation between the newly inserted syscall numbers in each level and the original syscall number in the trace. In $L1$, the increase percentage generally remains still, with a slight drop as the syscall number in the trace grows to 16000. In $L2$ and $L3$, the increased syscall number first drops when the trace syscall number grows to around 18000. Then, the success rate increases with the trace syscall number increasing. On average, in both $L1$, $L2$ and $L3$, the number of the increased syscalls does not have a relation with the syscall number in the trace.

\begin{table*}[!ht]
\centering
\caption{Statistics of the syscalls in the trace are indicated by the syscall type count. The total number of two successive dependent syscalls is denoted by the Successive Sequence (\#). The Max Children (\#) represents the maximum number of dependent subsequent syscalls that one syscall can have, and the Avg. Children (\#) denotes the average number of subsequent dependent syscalls.}
\label{tab:sequence_statstic}
\footnotesize
\begin{tabular}{c|c|c|c|c}
\hline
  Application  & Syscall type count & Successive Sequence (\#) & Max Children (\#)&Avg. Children (\#) \\
 \hline
 AdaptWatch&249&151&9&2.031\\
 DxDiag&154&70&12&1.875\\
 WordPad&311&189&7&1.947\\
 NotePad&250&152&10&1.862\\
\hline
Avg. & 241 & 140 & 9.5 & 1.929\\
\hline
\end{tabular}
\end{table*}
Table \ref{tab:sequence_statstic} shows the statistics of the syscalls in the trace. Even though the total syscall number is large, as in  \autoref{tab:seed_app}, the total types of syscalls is much less. Note that if a trace contains a syscall that is repeatedly invoked N times, then we have N syscall numbers while only 1 syscall type. All four seed apps' syscall type count are between 154 to 311 types. Additionally, note that not all types of syscalls can have subsequent dependent syscalls (the third column \texttt{Successive Sequence (\#)}). Approximately, only half of the syscall types have subsequent dependent syscalls in the trace. These are the dependent syscalls that \name learns from to generate the dictionary $\mathcal{D}$. For each such syscall, it can have at least one subsequent dependent syscall to a maximum of nearly 10 subsequent dependent syscalls (column \texttt{Max Children (\#)} of \autoref{tab:sequence_statstic}). On average, each such syscall can have 1.8 to 2.1 subsequent dependent syscalls (column \texttt{Avg. Children (\#)} of \autoref{tab:sequence_statstic}).


\subsection{Efficiency of \name}
\label{sec:exp_efficiency}

\begin{table}[!ht]
\centering
\caption{Efficiency of \name in seconds (\textbf{s}) on average. Each column corresponds to each step in \autoref{sec:meth}. Trace stands for Dynamic Syscall Tracing. Recover stands for Recover Syscall In The Trace. Synthesize stands for Synthesize Syscall Scrip. Mutate stands for Mutate Arguments.}
\label{tab:efficiency_offline}
\footnotesize
\begin{tabular}{c|c|c|c|c|c}
\hline
         Application &Trace &Dependency&Recover&Synthesize&Mutate\\
 \hline
 AdaptWatch&602&59&34&32&99\\
 DxDiag&597&61&37&28&96\\
 WordPad&620&67&37&35&110\\
 NotePad&580&49&29&28&86\\
 \hline
 Avg.&600&59&34&31&98\\
\hline
\end{tabular}
\end{table}

In this section, we test the efficiency of \name. The efficiency refers to two aspects: offline efficiency and online efficiency. Offline efficiency measures the time cost for each step in the methodology, including dynamic syscall tracing, syscall dependency analysis, model syscall script recovery, and syscall script synthesizing. Note that we do not measure the time for the last step, fuzzing, in the methodology, as fuzzing can take an infinite amount of time. Online efficiency refers to the average number of syscalls that can be executed per second.

\autoref{tab:efficiency_offline} shows the average result of offline efficiency. We averaged the results of all four seed apps. Generally, tracing takes the most time (approximately 10 minutes). The time delay comes mainly from two sources: 1) User interaction with the seed app. As we utilized the fuzzing module in NTfuzz for tracing, which prepared customized scripts to mimic user interaction with the apps, such as clicking buttons in the user GUI and striking keyboards, etc. To ensure each emulated user behaviour is correctly performed, it waits for some time (generally 1 second for each behaviour) after each behaviour. 
2) Writing the logs into the file. As we expanded the fuzzing module, the expanded fuzzing module writes the intercepted syscalls into the log file. This slows down the execution of the seed apps. For example, without logging, opening WordPad instantly pops up the user GUI (i.e., the WordPad Window) without any delay. However, with logging, Wordpad hangs for around 50 seconds before the window pops up. During this time, the trace logs are written to the file on the system. All other steps of \name take reasonably low time costs, varying between 30 seconds to 2 minutes.

\begin{table}[!ht]
\centering
\caption{The efficiency of \name's fuzzing is recorded in Column 2 (one iteration) as the average number of system calls per second, and Column 3 (multiple iteration) records the average number of system calls per second when taking initialization and waiting time into consideration.}
\label{tab:efficiency_online}
\footnotesize
\begin{tabular}{c|c|c}
\hline
         Application & Instant Rate (\# calls/s) & Averaged Rate (\# calls/s)\\
 \hline
 AdaptWatch&3770 &160 \\
 DxDiag&3697 &149\\
 WordPad&3744 &153\\
 NotePad&3756 & 158\\
 \hline
 Avg.& 3742 & 155\\
\hline
\end{tabular}
\end{table}

\autoref{tab:efficiency_online} shows the online fuzzing efficiency. With argument mutations, it is possible to cause Python errors or hanging errors that hinder the script from completing execution. Therefore, in the iterative fuzzing steps, we set a time limit for each iteration to ensure that the iteration can reach the next loop even when the current round encounters errors. \name achieves this by ignoring Python errors or hangings and continuing to the next fuzzing iteration. As shown in the table, the average instant fuzzing rate is 3742 system calls per second. However, since we automatically fuzz the script iteratively, extra time can be spent on 1) compiling the .py script in each iteration, and 2) the waiting time limit in the iteration that we set to ignore Python errors and hangings. After we average these extra time costs, the average fuzzing rate drops to around 155 calls per second.

We synthesized the syscall script into the python script because the Ctypes package makes it easy to call Windows syscalls. However, the python script needs to be compiled each time before it can be executed, and if the script contains a large number of syscalls, the compilation process can take a long time. Additionally, during fuzzing, errors in the syscall from the script can cause the fuzzing to be interrupted. These errors are usually caused by NtFuzz incorrectly inferring type information and the mismatch between the inferred argument and the actual usage in the trace, which can lead to the wrong argument values or types being used when calling the syscall. 

For example, illegal input argument values or insufficient memory allocation can cause errors. These cases can result in execution interruption with Python hangs or error messages. During our experiments, we deleted the syscalls belonging to the former case as we could not know the correct argument value. We allocated more space for the memory for the latter case.

\subsection{Crashes Found During Fuzzing}
\label{sec:exp_crash}
\begin{figure*}[!t]
\begin{subfigure}[b]{0.375\textwidth}
         \centering
         \includegraphics[width=\textwidth]{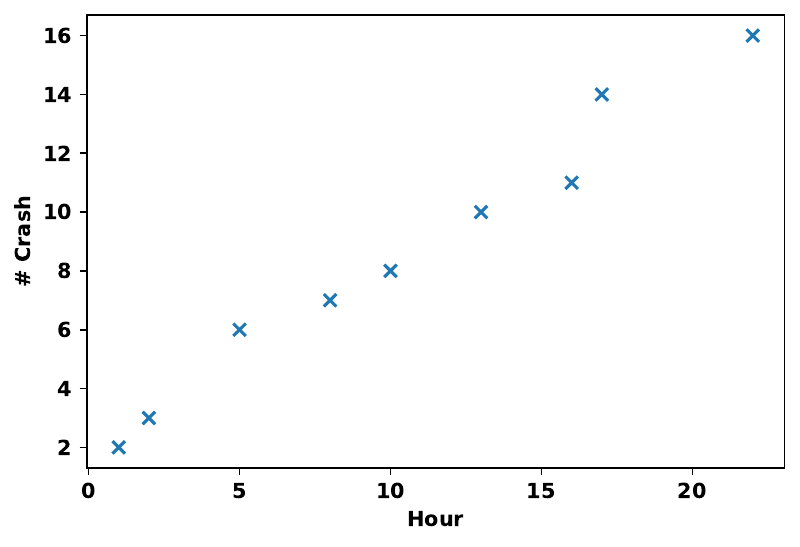}
         \caption{AdaptWatch}
         \label{fig:crash_awatch}
\end{subfigure}
\begin{subfigure}[b]{0.375\textwidth}
         \centering
         \includegraphics[width=\textwidth]{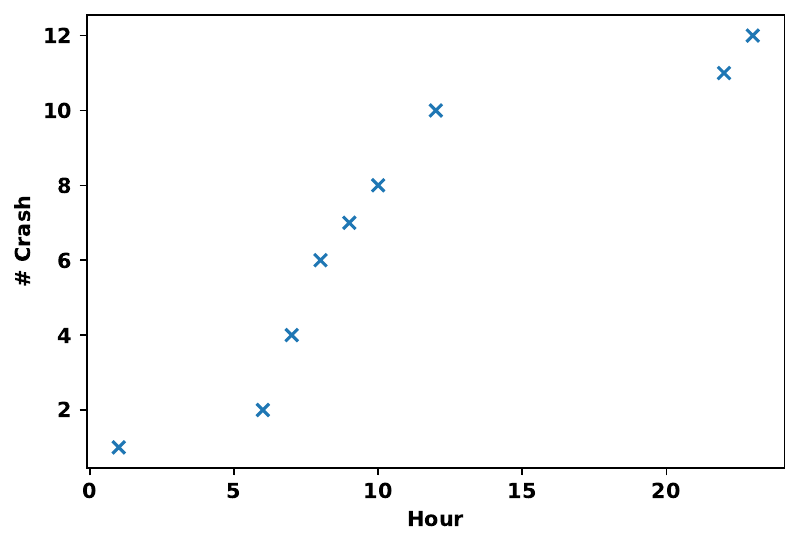}
         \caption{DxDiag}
         \label{fig:crash_dxdiag}
\end{subfigure}

\begin{subfigure}[b]{0.375\textwidth}
         \centering
         \includegraphics[width=\textwidth]{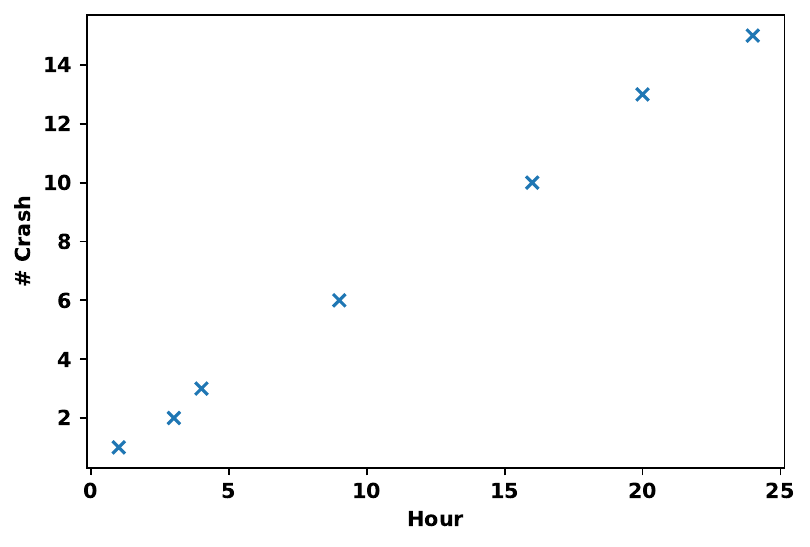}
         \caption{WordPad}
         \label{fig:crash_wordpad}
\end{subfigure}
\begin{subfigure}[b]{0.375\textwidth}
         \centering
         \includegraphics[width=\textwidth]{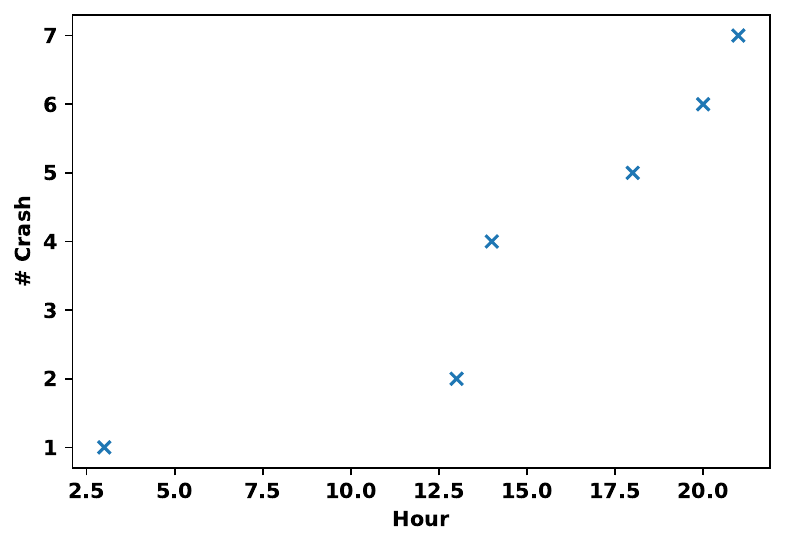}
         \caption{NotePad}
         \label{fig:crash_notepad}
\end{subfigure}

\caption{Number of crash versus time. We run each seed app with 24 hours.}
\label{fig:crash}
\end{figure*}

We are now assessing whether fuzzing the synthesized script can cause Windows systems to crash during fuzzing. If the Windows system crashes, it will automatically reboot and create a memory dump, or it may hang indefinitely. We fuzzed the synthesized scripts of the four seed applications, each with a 24-hour time limit. The results are shown in \autoref{fig:crash}. 

The general trend is that the number of crashes increases with time, even though sometimes there are no more crashes found within a relatively long time period (e.g., from hour 13 to 22 in DxDiag, and from hour 3 to 12 in NotePad). AdaptWatcher had the most crashes within the time frame, with 16 in total. WordPad and DxDiag followed with 15 and 12 crashes respectively. NotePad had the least total number of crashes, with 7 in total. We suspect this is because 1) NotePad had the least total number of syscalls with 7487 among the four seed apps, much lower than the other apps, as shown in \autoref{tab:seed_app}. This results in less probability for crashes to occur. 2) The types of syscalls called are different from the other seed applications, which may make it less likely to crash. To understand the syscall differences, in \autoref{tab:crash_syscall_statstics} we compare the statistics of the top 10 syscalls made by each seed app, along with the total times each syscall type was called in the trace. Since AdaptWatch, DxDiag, and WordPad have similar total crash numbers, we mainly compare the syscall difference between NotePad and the other three apps. We note that font-related (e.g., NtGdiGetFontData, NtGdiGetDCDword) and graphic-related syscalls (e.g., NtGdiSelectBitmap, NtUserSetWindowLong), and some other syscalls are called less in NotePad. We suspect the syscall type differences in NotePad also lead to fewer crashes.

\begin{table*}[!ht]
\centering
\caption{Top 10 most frequent syscalls in each seed app with the total time they are called during tracing.}
\label{tab:crash_syscall_statstics}
\footnotesize
\begin{tabular}{c|c|c|c|c|c|c|c|c}
\hline
    &Adaptwatch &\# & DxDiag &\# &WordPad &\# &NotePad &\# \\
 \hline
 1 & NtUserGetProp & 4186 & NtReadVirtualMemory & 2442 &NtGdiGetFontData & 2445 & NtQueryVirtualMemory & 1552\\
 \hline
2
&NtQueryValueKey & 904
&NtClose & 1894
&NtClose& 1984
&NtClose& 280\\
\hline
3
&NtGdiGetDCDword& 762
&NtQueryVirtualMemory& 1447
&NtQueryInformationProcess& 1828
&NtQueryValueKey& 273\\
\hline
4
&NtClose& 567&
NtOpenKeyEx& 1317&
NtUserGetProp& 1725&
NtOpenKeyEx& 243\\
\hline
5
&NtUserIsTopLevelWindow& 415&
NtQueryValueKey& 1128&
NtUserGetDpiForCurrentProcess& 1442&
NtUserGetProp& 225\\
\hline

\multirow{2}{*}{6}
&\multirow{2}{*}{NtUserSetWindowLong}& \multirow{2}{*}{397}
&\multirow{2}{*}{NtWaitForMultipleObjects}& \multirow{2}{*}{1012}
&\multirow{2}{*}{NtOpenKeyEx}& \multirow{2}{*}{1396}&
NtUserIsChildWindow-& \multirow{2}{*}{217}\\
&&&&&&&DpiMessageEnabled&\\
\hline
7
&NtGdiIntersectClipRect& 297&
NtQueryInformationToken& 1006&
NtGdiSelectBitmap& 1359&
NtQueryInformationToken& 203\\
\hline
8
&NtGdiGetRandomRgn& 292&
NtQueryKey& 830&
NtGdiDeleteObjectApp& 1334&
NtUserIsTopLevelWindow& 155\\
\hline

\multirow{2}{*}{9}
&NtUserIsChildWindow-&\multirow{2}{*}{284}&
\multirow{2}{*}{NtWaitForSingleObject}& \multirow{2}{*}{547}&
\multirow{2}{*}{NtUserSystemParametersInfo}& \multirow{2}{*}{1297}&
\multirow{2}{*}{NtTraceControl}& \multirow{2}{*}{148}\\

&DpiMessageEnabled &&&&&&&\\
\hline
10
&NtUserMessageCall& 273
&NtQueryInformationProcess& 292&
NtQueryValueKey& 1157&
NtEnumerateValueKey& 134\\

\hline
\end{tabular}
\end{table*}
\section{Discussion}
\mypara{Script Error Locating}
In the experiment, we manually changed the generated code to fix problems when calling syscalls from the generated scripts. Argument type errors can cause fatal kernel errors such as insufficient memory. These errors can cause the generated syscall sequence to get stuck in the middle of execution, leaving the rest of the syscalls not executed. It is difficult for humans to locate the error syscall number because the debugger can only tell the user that python has stopped working, not the exact location of the error. This requires humans to manually find the error in the script, which can contain tens of thousands of syscalls. Automatically locating the erroneous generated syscalls sites is a promising future direction to help increase efficiency. A possible solution is to create a tool to automatically comment out syscalls in the script iteratively to locate the erroneous syscall argument.

\mypara{Severity Prediction}
Even though we select seed applications randomly, it is possible that some applications can lead to a larger amount or higher severity of kernel vulnerabilities than others. As we have shown in \autoref{sec:exp_crash}, NotePad can lead to fewer system crashes due to its fewer number of syscalls and the different types of syscalls compared to other seed apps. Therefore, blindly selecting random seed applications might not be an efficient solution to find valuable kernel vulnerabilities. Therefore, static analysis and vulnerability severity prediction is also another direction that requires exploring. Further work can learn from large scale seed applications and their known vulnerabilities. An interesting direction is to find out the syscall types or sequences that are more likely to cause vulnerabilities through fuzzing. The learnt knowledge can be used to find applications that have a higher probability of causing crashes. Therefore, prior to fuzzing, one can statically analyze and filter the seed applications with a higher probability of causing system crashes.

\mypara{Input Optimization}
Unlike the Linux kernel, the Windows kernel is closed-source. This makes it difficult to test whether new fuzzing approaches can provide more optimized outcomes, such as increased code coverage, which is considered as feedback to guide the mutation step as it is assumed that more code coverage generally leads to more crashes. Rebert et al. \cite{rebert2014optimizing} optimized interesting input, while Woo et al. \cite{woo2013scheduling} and Cha et al. \cite{cha2015program} used a similar approach to optimize mutation rate. Godefroid et al. \cite{godefroid2012sage,godefroid2005dart}, Cadar et al. \cite{cadar2008klee}, Molnar et al. \cite{molnar2009dynamic}, Cha et al. \cite{cha2012unleashing}, Wang et al. \cite{wang2010taintscope}, Ganesh et al. \cite{ganesh2009taint}, and Rawat et al. \cite{rawat2017vuzzer} found interesting input by exploring new inputs that can reach deep and hidden code paths. However, it is impractical to directly apply the strategies from Linux or user-level application (i.e., adding codes in the kernel or application to probe the change of the code coverage) on Windows systems. Without this feedback information, fuzzing is not guided and is blindly mutated. Therefore, a novel approach that provides feedback information to guide fuzzing can be greatly beneficial to Windows kernel fuzzing. One possible solution is the kAFL \cite{kafl} approach, which is based on hypervisor and Intel’s Processor Trace (PT) technology. Combining the techniques from kAFL to guide the mutation direction is a promising direction.

\mypara{Script Synthesis}
Although our current implementation in Python is more efficient for calling Windows syscalls, it has been demonstrated in \autoref{sec:exp_efficiency} that calling the Python script iteratively slows down the fuzzing step significantly. This is due to the time taken for the script to compile and the waiting time in each iteration. We attempted to reduce the compilation time by generating executable files instead, using tools such as pyinstaller and cython. However, these tools claim to be able to transform Python scripts into executable programs (.exe files), but we were unable to produce any working programs due to memory errors caused by the large number of syscalls we synthesized. Therefore, it is possible to generate executable files from other languages, such as C or C++, instead of generating Python scripts, from the beginning. However, these languages do not provide packages that can be used to directly call Windows syscalls as in Python. It is still possible to implement the program as suggested in \cite{call_syscall_with_c}, but this would require more engineering effort.

\mypara{Syscall Dependency} Our approach provides a method to analyze system call dependencies by observing the input and output values from the trace. However, this method may have false positives. To increase accuracy, multiple traces can be logged and their inferred dependency information can be aggregated to reach a final conclusion. \cite{bastani2017synthesizing, godefroid2017learn,viide2008experiences} provide hints on solving this problem by using program execution traces to infer the whole input grammar. Moreover, Syzgen \cite{syzgen} uses symbolic execution to analyze macOS system calls from driver binaries. Even though this approach cannot be directly applied to Windows (as our target Windows system calls can only be accessed dynamically by using WinDbg and not any binaries), extracting Windows kernel routines (functions) and analyzing them symbolically offline is a promising direction to find Windows system call dependencies. Another practical method to learn dependencies among system calls is based on NtFuzz's static analysis of system binaries. Instead of voting for the majority system call argument type, all observed symbolized system call argument types can be preserved. Matching the symbolic expression between different system calls' arguments can potentially find more dependencies that are not logged in the trace. Therefore, we can use this statically learnt knowledge to generate new system call sequences, thus resulting in new kernel states.
\section{related work}
\mypara{Application fuzzing}
Langfuzz \cite{langfuzzing} learns the grammar of existing problematic testing cases (which lead to bugs or vulnerabilities) and combines and mutates the learnt code fragments to generate new testing code that could potentially lead to bugs or vulnerabilities. Langfuzz is a general approach that can be applied to different programming languages. Randoop \cite{Randoop} generates new valid unit testing cases (i.e., method sequences) based on feedback information to guide and optimize the generation. Randoop discards the generated cases that are invalid according to the feedback, and keeps the valid ones to generate new test cases.
WinAFL is a grey-box fuzzer that provides coverage feedback. It transplanted the approach of AFL from Linux user-level applications to Windows user-level applications. It mutates the input by observing and optimizing the feedback information (i.e., code coverage). However, current WinAFL implementations are unreliable in their persistent mode, which limits the applications it can support.
Winnie \cite{winnie} dynamically traces the application during execution and generates harness code based on the trace and applies process cloning to boost the fuzzing efficiency. The harness code can prevent the tested binary code from executing the GUI-based code each time they fuzz, which effectively reduces the time spent on not-interested code execution. However, this type of work is designed for tracing applications and cannot be directly applied to kernel objects. This reduces the time overhead to re-execute the GUI code every time they fuzz the application and boosts the fuzzing efficiency. However, this type of work targets user-level applications rather than the kernel. Moreover, Winnie cannot be directly used to synthesize harness code for kernel objects as they were designed for user-level executable or library files.

\mypara{Kernel fuzzing}
ReactOS \cite{ReactOS} provides some information about the argument types of system calls, but this information can be out-of-date. Most of the results are inferred manually. NTfuzz \cite{ntfuzz} fuzzes the Windows kernel by modifying the Windows system's SSDT. It runs randomly selected seed applications to trigger valid syscall sequences, as valid application execution always results in valid sequences. The modified SSDT table intercepts the syscall and randomly mutates the arguments to fuzz the kernel. NTfuzz also symbolically infers the syscall argument template to mutate the argument fields with more fine-grained and accurate granularity, which greatly reduces the time needed to find new bugs. \cite{model_fuzz} is a similar work to NtFuzz. It dynamically executes applications in Windows system and logs the executed syscalls to generate a model program that preserves the calling of the syscalls. Then it fuzzes the model program with mutated arguments. Moonshine \cite{moonshine} fuzzes the Linux kernel by tracing the system calls, then distilling them by identifying the dependencies between system calls. Syzgen \cite{syzgen} automatically generates syscall templates for macOS's driver by analyzing dependencies between driver binaries and iteratively refining the template. KAFL\cite{kafl} addresses the problem of kernel code coverage regardless of whether the system is open source or closed source. It applies a hypervisor and Intel’s Processor Trace (PT) technology to receive kernel level feedback information to guide the mutation direction. IMF \cite{imf} uses multiple logs of the program to infer the valid syscall ordering (syscall dependencies) with well-formed argument values that follow the API specification. Then it generates a model program based on the inferred information to fuzz the kernel iteratively. However, some techniques in this genre of work cannot be applied to closed-source systems like Windows. For example, Moonshine \cite{moonshine} requires modifying Linux source code, which is impractical for Windows and macOS. Other approaches generate model programs based on real program execution. The model program preserves the legitimate syscall ordering and dependencies, but the model program can be further mutated by learning dependencies from the trace. This only recovers the syscalls, as the trace can miss many corner cases that can potentially lead to different kernel states.  
\section{Conclusion}
To finish, we proposed a new way of fuzzing Windows kernel. Winkfuzz follows the syscalls that are called, along with their arguments and return values. It then looks at the connections between syscalls based on the trace and the argument type information from NTfuzz. Winkfuzz creates a model syscall script based on the traced syscall sequences and the identified syscall connections. After that, Winkfuzz adds syscalls to the model script to make new syscall sequences. Finally, Winkfuzz randomly mutates the argument field each time it runs the created syscall to fuzz the kernel.

\bibliographystyle{ACM-Reference-Format}
\bibliography{reference}

\end{document}